\documentclass[14pt, fleqn]{extarticle}

\usepackage{latexsym}
\usepackage{amsmath}
\usepackage[colorlinks = true,
            linkcolor = blue,
            urlcolor  = blue,
            citecolor = red,
            anchorcolor = blue]{hyperref}
\usepackage{amsfonts}
\usepackage{amssymb}
\usepackage{graphicx}
\usepackage{color}
\usepackage{tikz}
\usepgflibrary{arrows}
\usepackage{mathbbol}
\usepackage{xfrac}
\usepackage[a4paper,bindingoffset=0.2in,
           left=1in,right=1in,top=1in,bottom=1in]{geometry}

\renewcommand{\labelitemi}{$(i)$}
\renewcommand{\labelitemii}{$(ii)$}
\renewcommand{\labelitemiii}{$(iii)$}

\title{Global Symmetries of the Kepler Problem}
\author{
Joanna Gonera \footnote {joanna.gonera@uni.lodz.pl}, Piotr Kosi\'nski\footnote {piotr.kosinski@uni.lodz.pl}, Patryk Michel\footnote{michel.patryk@gmail.com}\\
\small \textit {Faculty of Physics and Applied Informatics}\\
\small \textit {University of Lodz, Poland}\\
}

\date{}
\begin{document}
\maketitle
\begin{abstract}
\par The global symmetry transformations generated by Runge-Lenz vector of twodimensional Kepler problem are explicitly described. They are given in terms of $ SU(2) $ left group multiplication with group elements being suitably parametrized by phase space points. The resulting nonlinear action of $SU(2)$ on the phase space is characterized in terms of the theory of nonlinear realizations developed in Phys. Rev. 177 (1969), 2239.
\end{abstract}

\section{Introduction} 
\label{I}
\par All static central potentials share four integrals of motion resulting from the symmetries under 
translations in time and rotations in space; these are energy and three components of angular momentum 
\cite{b1}. Three integrals, the length and one component of angular momentum and energy are in 
involution. Therefore, the dynamics is completely integrable \cite{b2}. One can still add another 
component of angular momentum as independent integral of motion so the system is superintegrable. 
However, in the generic case it is not maximally superintegrable; this would demand five independent 
integrals to exist. As a result most bounded trajectories are not closed but cover densely some ring in 
the plane of motion.
\par According to Bertrand's theorem \cite{b3}, \cite{b2} there are two noteworthy exceptions: the harmonic and Kepler potentials. In both cases we are dealing with maximally superintegrable systems. There are five functionally independent integrals of motion and all bounded orbits are closed. For the Kepler potential the additional integral$(s)$ is provided by Runge-Lenz vector \cite{b1}, \cite{b2}, \cite{b4,b5,b6}.
\par Both energy and angular momentum conservation result, via Noether theorem, from point symmetries, i.e. those defined, basically, in configuration space. This is no longer the case for additional integrals. In fact, they are related to more general symmetries described by canonical transformations \cite{b7,b8,b9}. Once this becomes clear it is quite straightforward to write out the infinitesimal canonical symmetry transformations leading, via the Hamiltonian form of Noether theorem, to conservation of Runge-Lenz vector. The natural question arises what is the global counterpart of such infinitesimal transformations. This problem has been studied in a number of papers \cite{b10,b11,b12,b13,b14,b15,b16,b17}. The success has been, however, slightly limited due to high nonlinearity the relevant equations.
\par In the present paper we provide more complete discussion of the global symmetry transformations for the Kepler problem. The equations determining infinitesimal transformations are viewed as Hamiltonian dynamics with the Hamiltonian being a linear combination of the components of Runge-Lenz vector. The resulting dynamics appears to be integrable and can be solved by Liouville-Arnold method \cite{b2}. We find the explicit solution in terms of action-angle variables defined for original Kepler dynamics. In particular, with properly normalized Runge-Lenz vector we find that the global symmetry transformations are simply given by left multiplication in the symmetry group (which is $SU(2)$ because we consider mainly $E<0$ and make use of the fact that the problem reduces to twodimensional one) properly parametrized in terms of action-angle variables. In this way we obtain a transparent geometric interpretation of global symmetry. We show how the global transformations can be conveniently described in terms of the method of nonlinear realizations developed in \cite{b19}. This is the main result of our paper. 
\par To conclude, let us note that in order to rewrite our results in terms of initial (Cartesian) coordinates and momenta we have to solve only one transcendental equation which is basically the Kepler equation. So we can say that the problem is solved explicitly to the same extent as the initial Kepler dynamics. 

\section{Noether theorem in Lagrangian\\ and Hamiltonian formalisms} 
\label{II}
\par The standard form of the Noether theorem refers to the Lagrangian formalism. Given a coordinate manifold and an action functional one considers the transformations of generalized coordinates (and, in general, time), called point transformations. Their main property is that the new coordinates (and time) depend only on the old ones and time; there is no dependence on generalized velocities. Assuming that the Lagrangian is form-invariant, up to a total time derivative of some function on configuration space, one can construct a conserved quantity, i.e. a function of coordinates, velocities and time which is constant along trajectories. Noether theorem gives an explicit form of the conserved charge. Basically, it is linear in generalized momenta; the nonlinear terms can enter only through the Hamiltonian provided time is also affected by the transformations under consideration. It is clear that, due to the specific form of conserved charge not all integrals of motion can be obtained in this way (even if they are globally well defined over the whole phase space). To see that this is the case it is sufficient to consider any nonlinear function of the integral of motion resulting from some point symmetry. It generates the canonical transformation which provides a symmetry on the Hamiltonian level.
\par The set of symmetry-related integrals of motion may be enlarged by referring to the Hamiltonian formalism. Then the Noether theorem can be generalized to the canonical transformations (i.e. those preserving the symplectic structure). Namely, let $G(q,p,t)$ be a generator of canonical transformations which infinitesimal form reads
\begin {align} 
\label {al1}
\delta q_{i}&=\delta\epsilon \{q_{i},G\}\\ \nonumber
\delta p_{i}&=\delta\epsilon \{p_{i},G\}
\end {align}
where $\delta\epsilon$ is an infinitesimal parameter such that $\delta\epsilon\cdot G$ has the dimension of action. Assuming that the Hamiltonian $H$ is form invariant under (\ref{al1}) one immediately obtains
\begin {align} 
\label {al2}
\{G,H\}+\frac{\partial G}{\partial t}=0
\end {align}
i.e. $G$ itself is a conserved charge. $G$ is, in general, nonlinear in momenta showing that the Hamiltonian version of the Noether theorem yields larger class of integrals of motion.\\
Eqs. (\ref{al1}) define an infinitesimal transformation of phase space. In order to recover their global form one has to solve the set of equations. 
\begin {align} 
\label {al3}
\frac{dq_{i}}{d\epsilon}&=\{q_{i},G\}\\ \nonumber
\frac{dp_{i}}{d\epsilon}&=\{p_{i},G\}
\end {align}
Eqs. (\ref{al3}) have the Hamiltonian form with the "Hamiltonian" $G$ and "time" $\epsilon$. Therefore, all standard techniques of Hamiltonian formalism (canonical transformations, action-angle variables, Liouville integrability, Hamiltonian-Jacobi equation) can be applied. In particular, for two degrees of freedom $(i=1,2)$ and $G$ does not depending explicitly on time there are two Poisson commuting "integrals of motion", $G$ and $H$, so eqs. (\ref{al3}) are Liouville integrable.

\section{The Kepler problem} 
\label{III}

\par The Kepler problem is the textbook example of analytically solvable Newtonian system. It is defined by the Hamiltonian
\begin {align} 
\label {al4}
H=\frac{\vec{p}\,^{2}}{2m}-\frac{k}{\vert\vec{q}\vert}
\end {align}
with $m$ and $k$ being the mass and coupling constant, respectively, while $\vec{q}\in\mathbb{R}^{3}$, $\vec{p}\in\mathbb{R}^{3}$.
\par The system exhibits two basic point symmetries: time translation, implying energy conservation $(H\equiv E=const.)$ and rotations, leading to the conservation of orbital angular momentum
\begin {align} 
\label {al5}
\vec{L}\equiv \vec{q}\times \vec{p}\quad \text{;}
\end {align}
these integrals are shared with all central conservative potentials. The corresponding Hamiltonians define Liouville integrable systems: as Poisson commuting integrals of motion one can take $H$, $L_{3}$ and $\vec{L}^{2}$.
\par However, for the specific potential (\ref{al4}) there exists an additional integral of motion called Runge-Lenz vector and defined by
\begin {align} 
\label {al6}
\vec{A}=\vec{p}\times\vec{L}-\frac{mk\vec{q}}{\vert\vec{q}\vert}
\end {align}
\par Therefore, we have seven integrals of motion: $H$, $\vec{L}$ and $\vec{A}$. They are not independent; in fact, one easily verifies the following identities
\begin {align} 
\label {al7}
\vec{A}\cdot \vec{L}=0
\end {align}
\begin {align} 
\label {al8}
\vec{A}\,^{2}=m^{2}k^{2}+2mE\vec{L}\,^{2}
\end {align}
\par So we are left with five independent integrals which is the maximal allowed number for systems with three degrees of freedom. Such a system is called maximally superintegrable.
\par The Poisson bracket of two integrals of motion is again an integral of motion. In the particular case of Kepler problem the Poisson algebra reads
\begin {align} 
\label {al9}
\{H,\vec{L}\}=\{H,\vec{A}\}=0
\end {align}
\begin {align} 
\label {al10}
\{L_{i}, L_{j}\}=\epsilon_{ijk}L_{k}
\end {align}
\begin {align} 
\label {al11}
\{L_{i}, A_{j}\}=\epsilon_{ijk}A_{k}
\end {align}
\begin {align} 
\label {al12}
\{A_{i},A_{j}\}=-2mH\epsilon_{ijk}L_{k}
\end {align}
It follows that $(\vec{L},\vec{A})$ form a Lie algebra on any submanifold $H=E$ of phase space. It is $sO(4)$, $e(3)$ or $sO(1,3)$ depending on whether $E<0$, $E=0$ or $E>0$, respectively.
\par Assuming $\vec{L}$ and $\vec{A}$ are known one easily finds the trajectories. First, the particle moves in a plane perpendicular to $\vec{L}$, $\vec{q}\perp\vec{L}$ and $\vec{p}\perp\vec{L}$. Rewriting (\ref{al6}) as
\begin {align} 
\label {al13}
\frac{mk\vec{q}}{\vert\vec{q}\vert}=\vec{p}\times \vec{L}-\vec{A}
\end {align}
and taking the square we find
\begin {align} 
\label {al14}
\Bigg (\vec{p}-\frac{\vec{L}\times\vec{A}}{\vec{L}\,^{2}}\Bigg)^{2}=\frac{m^{2}k^{2}}{\vec{L}\,^{2}}
\end {align}
Therefore, the hodograph, being the intersection of the sphere (\ref{al14}) with the plane $\vec{L}\cdot \vec{p}=0$, is a circle. Moreover, denoting by $\varphi$ the angle between $\vec{q}$ and $\vec{A}$, one finds by virtue of eq. (\ref{al6})
\begin {align} 
\label {al15}
\vert\vec{q}\vert=\frac{\frac{\vec{L}\,^{2}}{mk}}{1+\frac{\vert\vec{A}\vert}{mk}\cos \varphi}
\end {align}
which is the equation of conic section. 
\par One can simplify the description by taking explicitly into account that we are dealing with plane motion. Let the plane of motion be the $1 2$ plane. Then $\vec{q}\equiv (q_{1}, q_{2})$, $\vec{p}\equiv (p_{1}, p_{2})$ and $\vec{A}\equiv (A_{1}, A_{2})$ are twodimensional vectors while $L_{3}\equiv L$ represents orbital angular momentum. Eqs. (\ref{al5}), (\ref{al6}), (\ref{al11}), (\ref{al12}) read, respectively
\begin {align} 
\label {al16}
L=\epsilon_{ij}q_{i}p_{j}
\end {align}
\begin {align} 
\label {al17}
A_{i}=\epsilon_{ij}p_{j}L-\frac{mkq_{i}}{\vert\vec{q}\vert}
\end {align}
\begin {align} 
\label {al18}
\{L,A_{i}\}=\epsilon_{ij}A_{j}
\end {align}
\begin {align} 
\label {al19}
\{A_{i},A_{j}\}=-2mHL\epsilon_{ij}
\end {align}
\par We see that the initial symmetry algebra is "spontaneously" broken to $sU(2)$ (or, equivalently, $sO(3)$), $e(2)$, $sU(1,1)$, depending on whether $E<0$, $E=0$ or $E>0$, respectively. The hodograph equation (\ref{al14}) may be rewritten as
\begin {align} 
\label {al20}
\big(\vec{p}-\vec{D}\big)^{2}=\frac{m^{2}k^{2}}{L^{2}}
\end {align}
with
\begin {align} 
\label {al21}
D_{i}=-\frac{\epsilon_{ij}A_{j}}{L}
\end {align}
\par In what follows we always assume that our Cartesian coordinates are chosen as described above. So, basically, we are considering twodimensional Kepler problem; the results can be easily rewritten in threedimensional language. 

\section{Symmetry transformations generated by Runge-Lenz vector} 
\label{IV}
\par By inspecting the momentum dependence of the Runge-Lenz vector (\ref{al6}) one concludes that it cannot represent the conserved charge resulting from point symmetry. Therefore, in order to relate its conservation to some symmetry we have to refer to Noether theorem in the Hamiltonian form. According to the discussion in Sec. II we consider the canonical transformations generated by
\begin {align} 
\label {al22}
G=A_{c}\equiv \vec{c}\cdot\vec{A}
\end {align}
where $\vec{c}\equiv(c_{1},c_{2})$ is a constant vector which can be taken to be the unit one, $\vert\vec{c}\vert=1$, by an appropriate normalization of the evolution parameter $\epsilon$. $A_{c}$, being conserved, generates canonical symmetry transformations. The relevant set of "dynamical" equations, 
\begin {align} 
\label {al23}
\frac{dq_{i}}{d\epsilon}=\{q_{i},A_{c}\}
\end {align}
\begin {align} 
\label {al24}
\frac{dp_{i}}{d\epsilon}=\{p_{i},A_{c}\}
\end {align}
\begin {align} 
\label {al25}
A_{c}\equiv (\vec{c}\cdot\vec{q})\vec{p}\,^{2}-(\vec{c}\cdot\vec{p})(\vec{q}\cdot\vec{p})-\frac{mk(\vec{c}\cdot\vec{q})}{\vert\vec{q}\vert}
\end {align}
reads
\begin {align} 
\label {al26}
\frac{dq_{i}}{d\epsilon}=2(\vec{c}\cdot\vec{q})p_{i}-(\vec{c}\cdot\vec{p})q_{i}-(\vec{q}\cdot\vec{p})c_{i}
\end {align}
\begin {align} 
\label {al27}
\frac{dp_{i}}{d\epsilon}=-\vec{p}\,^{2}\cdot c_{i}+(\vec{c}\cdot\vec{p})p_{i}+\frac{mk}{\vert\vec{q}\vert}\Bigg (c_{i}-\frac{(\vec{c}\cdot\vec{q})q_{i}}{\vert\vec{q}\vert^{2}}\Bigg)
\end {align}
Eqs. (\ref{al26}), (\ref{al27}) constitute the set of four nonlinear differential equations and look quite complicated. However, they admit two Poisson commuting integrals of motion, the "Hamiltonian" $A_{c}$ and the initial Hamiltonian $H$, eq. (\ref{al4}). Therefore, the "dynamics" described by eqs. (\ref{al26}), (\ref{al27}) is Liouville integrable and can be solved by quadratures. 
\par Before describing the details of Liouville-Arnold procedure for eqs. (\ref{al26}), (\ref{al27}) we show how the whole problem of solving them can be reduced to that of solving a single nonlinear differential equation. To this end let us remind that the dynamics (\ref{al26}), (\ref{al27}) leaves the submanifold $H=E$ invariant. On this manifold $\vec{A}$ and $L$ span the Lie algebra with respect to the Poisson bracket. Therefore, the evolution equations for $\vec{A}$ and $L$ linearize. In what follows we shall consider the most interesting case $E<0$; the results can be, however, easily generalized to the case $E\geq 0$. Using the explicit form of Poisson brackets, eqs. (\ref{al18}), (\ref{al19}), one finds
\begin {align} 
\label {al28}
\frac{dL}{d\epsilon}=\epsilon_{ij}c_{i}A_{j}
\end {align}
\begin {align} 
\label {al29}
\frac{dA_{i}}{d\epsilon}=\omega^{2}L\epsilon_{ij}c_{j}\qquad\text {,}\qquad \omega^{2}\equiv-2mE=2m\vert E \vert
\end {align}
Their solution reads
\begin {align} 
\label {al30}
L(\epsilon)=L(0)\cos(\omega\epsilon)+\frac{\epsilon_{ij}c_{i}A_{j}(0)}{\omega}\sin(\omega\epsilon)
\end {align}
\begin {align} 
\label {al31}
A_{i}(\epsilon)&=\big(A_{i}(0)-c_{i}c_{j}A_{j}(0)\big)\cos(\omega\epsilon)+\omega\epsilon_{ij}c_{j}L(0)\sin(\omega\epsilon)\\ \nonumber
&+c_{i}c_{j}A_{j}(0)
\end {align}
Eqs. (\ref{al4}) and (\ref{al17}) allow us to express $\vec{q}$ in terms of $\vec{p}$, $\vec{A}$ and $H$,
\begin {align} 
\label {al32}
\vec{q}=\frac{\big(mH-\frac{\vec{p}\,^{2}}{2}\big)\vec{A}+(\vec{p}\cdot\vec{A})\vec{p}}{m^{2}H^{2}-\big(\frac{\vec{p}\,^{2}}{2}\big)^{2}}
\end {align}
Since $H(\epsilon)=H\equiv E$ and $\vec{A}(\epsilon)$ (cf. eq. (\ref{al31})) are known, $\vec{q}(\epsilon)$ is determined uniquely once $\vec{p}(\epsilon)$ is known. In order to find $\vec{p}(\epsilon)$ we have only to insert $\vec{q}$, as given by eq. (\ref{al32}), into eq. (\ref{al27}). However, it is convenient to introduce first the suitable parametrization of $\vec{p}$. To this end we use eq. (\ref{al20}) and put
\begin {align} 
\label {al33}
\vec{p}=\vec{D}+\frac{mk}{L}\vec{n}
\end {align}
where
\begin {align} 
\label {al34}
\vec{n}\equiv (\cos \theta, \sin  \theta)
\end {align}
is a unit vector. Note that by virtue of eqs. (\ref{al21}), (\ref{al30}), (\ref{al31}) both $\vec{D}(\epsilon)$ and $L(\epsilon)$ are explicitly known. Therefore, it remains to determine $\theta (\epsilon)$. To this end we insert eqs. (\ref{al32}) and (\ref{al33}) into eqs. (\ref{al27}). This is simplified by noting that $\vec{q}\cdot \vec{n}=0$ (as follows from eq. (\ref{al32})) and the vector $\frac{\vec{q}}{\vert \vec{q}\vert}$ enters (\ref{al27}) only quadratically. The final result reads
\begin {align} 
\label {al35}
\tilde{\vec{n}}\cdot\frac{d\vec{n}}{d\epsilon}=-\frac{2mk}{L}(\vec{c}\cdot \tilde{\vec{n}})-2(\vec{n}\cdot\vec{D})(\tilde{\vec{n}}\cdot\vec{c})+(\vec{n}\cdot\vec{c})(\tilde{\vec{n}}\cdot \vec{n})
\end {align}
where
\begin {align} 
\label {al36}
\tilde{n}_{i}=\epsilon_{ij}n_{j}
\end {align}
i.e.
\begin {align} 
\label {al37}
\tilde{\vec{n}}=(\sin \theta,-\cos\theta)
\end {align}
Eq. (\ref{al35}) is the first order nonlinear differential equation for $\theta(\epsilon)$ with variable coefficients. In general, we do not expect such an equation to be integrable in quadratures, even in implicit form. However, as we have noted above, our "dynamics" is integrable. Consequently, eq. (\ref{al35}) should be solvable by quadratures. Once it is solved one can reconstruct $\vec{p}(\epsilon)$ from eq. (\ref{al33}) and, finally, $\vec{q}(\epsilon)$ from eq. (\ref{al32}). The whole procedure reduces, therefore, to solving linear dynamics on $sU(2)$ algebra (eqs. (\ref{al28}), (\ref{al29})) and single nonlinear eq. (\ref{al35}).
\par Instead of following this way we prefer to implement the more transparent Liouville-Arnold algorithm. First, we pass to polar coordinates (the  $L-A$ approach in Cartesian coordinates is sketched in Appendix)
\begin {align} 
\label {al38}
q_{1}&=r\cos\varphi\\ \nonumber
q_{2}&=r\sin\varphi
\end {align}
\begin {align} 
\label {al39}
p_{1}&=p_{r}\cos\varphi-\frac{p_{\varphi}}{r}\sin\varphi  \\ \nonumber
p_{2}&=p_{r}\sin\varphi+\frac{p_{\varphi}}{r}\cos\varphi 
\end {align}
Then
\begin {align} 
\label {al40}
H=\frac{p^{2}_{r}}{2m}+\frac{p^{2}_{\varphi}}{2mr^{2}}-\frac{k}{r}
\end {align}
and
\begin {align} 
\label {al41}
A_{1}=\Bigg(\frac{p^{2}_{\varphi}}{r}-mk\Bigg)\cos\varphi+p_{r}p_{\varphi}\sin\varphi\\\nonumber
A_{2}=\Bigg(\frac{p^{2}_{\varphi}}{r}-mk\Bigg)\sin\varphi-p_{r}p_{\varphi}\cos\varphi
\end {align}
Next, we introduce the action-angle variables $I_{r}$, $I_{\varphi}$, $\Theta_{r}$, $\Theta_{\varphi}$ for the original Kepler dynamics. Starting from the Poisson commuting integrals of motion $H$ and $p_{\varphi}$ we find
\begin {align} 
\label {al42}
I_{\varphi}=p_{\varphi}
\end {align}
\begin {align} 
\label {al43}
H=-\frac{mk^{2}}{2(I_{r}+I_{\varphi})^{2}}
\end {align}
\begin {align} 
\label {al44}
\Theta_{r}=-\sqrt{2mkr-\frac{m^{2}k^{2}r^{2}}{(I_{r}+I_{\varphi})^{2}}-I^{2}_{\varphi}}-\arcsin\left(\frac{1-\frac{mkr}{(I_{r}+I_{\varphi})^{2}}}{\sqrt{1-\frac{I^{2}_{\varphi}}{(I_{r}+I_{\varphi})^{2}}}}\right)
\end {align}
\begin {align} 
\label {al45}
\Theta_{\varphi}-\Theta_{r}=\varphi-\arcsin\left(\frac{1-\frac{I^{2}_{\varphi}}{mkr}}{\sqrt{1-\frac{I^{2}_{\varphi}}{(I_{r}+I_{\varphi})^{2}}}}\right)
\end {align}
\par Inserting on the left hand side of (\ref{al44}) $\Theta_{r}$ as linear function of time one obtains Kepler equation determining the time dependence of radial coordinate.
\par Now, using eqs. (\ref{al42})-(\ref{al45}) one can express the Runge-Lenz vector in terms of action and angle variables,
\begin {align} 
\label {al46}
A_{1}=mk\sqrt{1-\frac{I^{2}_{\varphi}}{(I_{r}+I_{\varphi})^{2}}}\sin(\Theta_{\varphi}-\Theta_{r})
\end {align}
\begin {align} 
\label {al47}
A_{2}=-mk\sqrt{1-\frac{I^{2}_{\varphi}}{(I_{r}+I_{\varphi})^{2}}}\cos(\Theta_{\varphi}-\Theta_{r})
\end {align}
In order to simplify notation we make further canonical transformation
\begin {align} 
\label {al48}
\Theta_{1}&=\Theta_{\varphi}-\Theta_{r}\\\nonumber
\Theta_{2}&=\Theta_{r}
\end {align} 
\begin {align} 
\label {al49}
I_{1}&=I_{\varphi}\qquad \qquad \text {,}\quad I_{2}\geq I_{1}\geq 0 \\ \nonumber
I_{2}&=I_{r}+I_{\varphi}
\end {align}
New angles also provide correct parametrization of Liouville-Arnold torus. In terms of new variables we find
\begin {align} 
\label {al50}
A_{1}=mk\sqrt{1-\frac{I^{2}_{1}}{I^{2}_{2}}}\sin\Theta_{1}
\end {align}
\begin {align} 
\label {al51}
A_{2}=-mk\sqrt{1-\frac{I^{2}_{1}}{I^{2}_{2}}}\cos\Theta_{1}
\end {align}
In order to compute $A_{c}$ let us put $\vec{c}=(\cos\chi, \sin\chi)$; then
\begin {align} 
\label {al52}
A_{c}=mk\sqrt{1-\frac{I^{2}_{1}}{I^{2}_{2}}}\sin(\Theta_{1}-\chi)
\end {align}
\par The replacement $\Theta_{1}\to\Theta_{1}-\chi$, with the remaining variables unchanged, defines the canonical transformation. Therefore, it is sufficient to consider the special case $\chi=0$, i.e. $A_{c}=A_{1}$; the general situation is recovered by replacing $\Theta_{1}$ by $\Theta_{1}-\chi$.
\par In terms of new variables the dynamics generated by $A_{1}$ is much simpler (let us stress again that our basic variables $\Theta_{i}$, $I_{i}$, $i=1,2$ are the action-angle variables for the \underline{original Kepler dynamics}). Therefore, the Liouville-Arnold method is available in explicit form. The Poisson-commuting integrals of motion, $A_{1}$ and $I_{2}$, serve as new momenta. We have to find the canonically conjugated variables $\Psi_{1}$, $\Psi_{2}$. To this end we express $I_{1}$ in terms of $\Theta_{1}$, $A_{1}$ and $I_{2}$ and compute the generating function
\begin {align} 
\label {al53}
S=\int\big(I_{1}\left(A_{1}, I_{2}, \Theta_{1}\right)d\Theta_{1}+I_{2}d\Theta_{2}\big)
\end {align}
The variables $\Psi_{1}$, $\Psi_{2}$ are defined as 
\begin {align} 
\label {al54}
\Psi_{1}=\frac{\partial S}{\partial A_{1}}\qquad \text{,}\qquad \Psi_{2}=\frac{\partial S}{\partial I_{2}}
\end {align}
Performing the relevant computations we find from eqs. (\ref{al50}), (\ref{al53}) and (\ref{al54}): 
\begin {align} 
\label {al55}
\Psi_{1}&=\frac{I_{2}}{2mk}\Bigg(\arcsin\left(\frac{\cos \Theta_{1}+1-a}{\sqrt{1-a}(1+\cos\Theta_{1})}\right)\\ \nonumber
&+\arcsin \left(\frac{\cos\Theta_{1}-1+a}{\sqrt{1-a}(1-\cos\Theta_{1})}\right)\Bigg)
\end {align}
\begin {align} 
\label {al56}
\Psi_{2}=\frac{mk\sqrt{a}}{I_{2}}\Psi_{1}-\arcsin\left(\frac{\cos\Theta_{1}}{\sqrt{1-a}}\right)+\Theta_{2}
\end {align}
where
\begin {align} 
\label {al57}
a\equiv\frac{A^{2}_{1}}{m^{2}k^{2}}<1
\end {align}
Due to the form of the "Hamiltonian" $(=A_{1})$ we find
\begin {align} 
\label {al58}
\frac{d\Psi_{1}}{d\epsilon}=1\qquad \text{,}\qquad \frac{d\Psi_{2}}{d\epsilon}=0
\end {align}
First eq. (\ref{al58}) together with eq. (\ref{al55}) implies 
\begin {align} 
\label {al59}
\sin^{2}\Theta_{1}(\epsilon)=\frac{2a}{(1+a)-(1-a)\cos\left(\frac{2mk\epsilon}{I_{2}}+\sigma_{1}\right)}
\end {align}
with $\sigma_{1}$ being an arbitrary constant to be determined from initial conditions. Note that due to the eq. (\ref{al50}) the sign of $\sin\Theta_{1}(\epsilon)$ is determined. Having $\Theta_{1}(\epsilon)$ determined we can compute $I_{1}(\epsilon)$ from eq. (\ref{al50}) $(I_{1}(\epsilon)\geq 0)$. On the other hand, $I_{2}(\epsilon)=I_{2}$ is a constant of motion so only $\Theta_{2}(\epsilon)$ must be determined. Eq. (\ref{al56}) yields
\begin {align} 
\label {al60}
\Theta_{2}(\epsilon)=\arcsin\left(\frac{\cos\Theta_{1}(\epsilon)}{\sqrt{1-a}}\right)-\frac{mk\sqrt{a}}{I_{2}}\epsilon+\sigma_{2}
\end {align}
with $\sigma_{2}$ being again an arbitrary constant to be determined from initial conditions. Once $\Theta_{i}(\epsilon)$, $I_{i}(\epsilon)$, $i=1,2$, are known, one compute $\Theta_{\varphi}(\epsilon)$, $\Theta_{r}(\epsilon)$, $I_{r}(\epsilon)$ and then, using (\ref{al42})-(\ref{al45}), the original phase space variables $r(\epsilon)$, $\varphi(\epsilon)$, $p_{r}(\epsilon)$, $p_{\varphi}(\epsilon)$. Note that the only transcendental equation which appears in the process is of the Kepler type. 
\par Although we have made a special choice $G=A_{1}$, the general case $G=A_{c}$ is obtained by making the canonical transformation $\Theta_{1}\to\Theta_{1}-\chi$.
\par Let us note that the trajectories described by eqs. (\ref{al50}), (\ref{al59}) and (\ref{al60}) are, in general, not closed. They belong to the submanifolds $H=const$ on which $\vec{A}$ and $\vec{L}$ close to $sU(2)$ algebra (with respect to Poisson brackets). The infinitesimal action of this algebra should be, by  Lie-Palais theorem, integrable to global $SU(2)$  action and all relevant trajectories closed. We discuss this point in the next section.

\section{The $SU(2)$ action} 
\label{V}
\par The symmetry transformations generated by the components of Runge-Lenz vector leave the submanifold $H=E$ invariant. Moreover, $\vec{A}$ and $L$ form $sU(2)$ algebra on this submanifold. However, the infinitesimal transformations generated by $\vec{A}$ and $L$ do not integrate to the global ones, because, generically, the relevant trajectories in phase space are not closed. The reason underlying such a behaviour is quite simple \cite{b18}. For any generator $G(\underline{q}, \underline{p})$ we define the corresponding vector field on phase space by
\begin {align} 
\label {al61}
X_{G}=-\{q_{i},G\}\frac{\partial}{\partial q_{i}}-\{p_{i},G\}\frac{\partial}{\partial p_{i}}
\end {align}
Then the following relation holds
\begin {align} 
\label {al62}
[X_{G},X_{G'}]=X_{\{G,G'\}}
\end {align}
Therefore, the basic Poisson bracket (\ref{al19}) implies
\begin {align} 
\label {al63}
[X_{A_{i}},X_{A_{j}}]=-2m\epsilon_{ij}(LX_{H}+HX_{L})
\end {align}
So, even on the submanifold $H=E$ we are not dealing with Lie algebra of vector fields. 
\par In order to improve this one can appropriately normalize the Runge-Lenz vector by defining (for $H<0$)
\begin {align} 
\label {al64}
B_{i}\equiv \frac{A_{i}}{\sqrt{-2mH}}\qquad i=1,2
\end {align}
\begin {align} 
\label {al65}
B_{3}\equiv L
\end {align}
Then $\{B_{i}\}$ span $sU(2)$ algebra in the standard basis
\begin {align} 
\label {al66}
\{B_{i},B_{j}\}=\epsilon_{ijk}B_{k}
\end {align}
so the corresponding infinitesimal canonical transformations should be integrable to global $SU(2)$ ones. $B'_{i}s$ are conserved and we obtain the $SU(2)$ symmetry transformations for the Kepler problem.
\par In terms of action-angle variables $B'_{i}s$ read
\begin {align} 
\label {al67}
B_{1}=\sqrt{I^{2}_{r}+2I_{r}I_{\varphi}}\sin(\Theta_{\varphi}-\Theta_{r})
\end {align}
\begin {align} 
\label {al68}
B_{2}=-\sqrt{I^{2}_{r}+2I_{r}I_{\varphi}}\cos(\Theta_{\varphi}-\Theta_{r})
\end {align}
\begin {align} 
\label {al69}
B_{3}=I_{\varphi}
\end {align}
or
\begin {align} 
\label {al70}
B_{1}=\sqrt{I_{2}^{2}-I_{1}^{2}}\sin\Theta_{1}
\end {align}
\begin {align} 
\label {al71}
B_{2}=-\sqrt{I_{2}^{2}-I_{1}^{2}}\cos\Theta_{1}
\end {align}
\begin {align} 
\label {al72}
B_{3}=I_{1}
\end {align}
\par Instead of $A_{c}$ one considers the generator
\begin {align} 
\label {al73}
B_{c}=\sqrt{I^{2}_{2}-I^{2}_{1}}\sin(\Theta_{1}-\chi)
\end {align}
and again it is sufficient to consider the special case $\chi=0$. The evolution equations generated by $B_{1}$ define integrable Hamiltonian system with two Poisson commuting integrals $B_{1}$ and $I_{2}$. Following Liouville-Arnold algorithm as in previous section we find the variables canonically conjugated to $B_{1}$ and $I_{2}$ (which, for simplicity, we again denote by $\Psi_{1,2}$):\
\begin {align} 
\label {al74}
\Psi_{1}=\arcsin\left(\frac{\cos\Theta_{1}+1-\beta}{\sqrt{1-\beta}(1+\cos\Theta_{1})}\right)+\arcsin\left(\frac{\cos\Theta_{1}-1+\beta}{\sqrt{1-\beta}(1-\cos\Theta_{1})}\right)
\end {align}
\begin {align} 
\label {al75}
\Psi_{2}=\Theta_{2}-\arcsin\left(\frac{\cos\Theta_{1}}{\sqrt{1-\beta}}\right)\\ \nonumber
\text {with} \quad 0\leq\beta\equiv\frac{B^{2}_{1}}{I^{2}_{2}}<1\quad \text {.}
\end {align}
Eq. (\ref{al74}), together with the evolution equation
\begin {align} 
\label {al76}
\frac{d\Psi_{1}}{d\epsilon}=1
\end {align}
yields
\begin {align} 
\label {al77}
\sin^{2}\Theta_{1}=\frac{2\beta}{(1+\beta)-(1-\beta)\cos(\epsilon+\sigma_{1})}
\end {align}
with $\sigma_{1}$ being an arbitrary constant. Again the sign of $\sin\Theta_{1}(\epsilon)$ is fixed so eq. (\ref{al77}) allows us to determine $\Theta_{1}(\epsilon)$ and then, due to the fact that $B_{1}$ is an integral of motion, also $I_{1}(\epsilon)$. It remains to find $\Theta_{2}(\epsilon)$. Second evolution equation
\begin {align} 
\label {al78}
\frac{d\Psi_{2}}{d\epsilon}=0
\end {align}
implies
\begin {align} 
\label {al79}
\Theta_{2}(\epsilon)=\arcsin\left(\frac{\cos\Theta_{1}(\epsilon)}{\sqrt{1-\beta}}\right)+\sigma_{2}
\end {align}
\par Eqs. (\ref{al70}), (\ref{al77}) and (\ref{al79}) clearly show that all trajectories in phase space are closed. Actually, apart from $B_{1}$ and $I_{2}$ one finds that there exists an additional independent and globally defined integral of motion which can be written as
\begin {align} 
\label {al80}
C=I_{1}\sin\Theta_{1}\sin\Theta_{2}-I_{2}\cos\Theta_{1}\cos\Theta_{2}
\end {align}
Our evolution equations are therefore superintegrable.
\par Once the action of symmetry transformations on action-angle variables is known one can, in principle, recover their action on initial variables $\vec{q}$ and $\vec{p}$. This, however, amounts to solve transcendental equation of the Kepler type.
\par It remains to read off the explicit form of $SU(2)$ action on the phase space parametrized by action-angle variables $I_{1}$, $I_{2}$, $\Theta_{1}$, $\Theta_{2}$. Let us write out explicitly the evolution equations generated by $B_{1}$. They read
\begin {align} 
\label {al81}
\frac{d\Theta_{1}}{d\epsilon}=\frac{\partial B_{1}}{\partial I_{1}}=-\frac{I_{1}\sin\Theta_{1}}{\sqrt{I^{2}_{2}-I^{2}_{1}}}
\end {align}
\begin {align} 
\label {al82}
\frac{d\Theta_{2}}{d\epsilon}=\frac{\partial B_{1}}{\partial I_{2}}=\frac{I_{2}\sin\Theta_{1}}{\sqrt{I^{2}_{2}-I^{2}_{1}}}
\end {align}
\begin {align} 
\label {al83}
\frac{dI_{1}}{d\epsilon}=-\frac{\partial B_{1}}{\partial\Theta_{1}}=-\sqrt{I^{2}_{2}-I^{2}_{1}}\cos\Theta_{1}
\end {align}
\begin {align} 
\label {al84}
\frac{dI_{2}}{d\epsilon}=0
\end {align}
Note that eq. (\ref{al83}) does not imply that the sign of $I_{1}$ is preserved - see the remark in Sec. VI. 
\par We would like to give a clear grouptheoretical interpretation of eqs. (\ref{al81})-(\ref{al84}). First note, that due to the fact that $I_{2}$ is constant we have to determine the global $SU(2)$ action on the $I_{2}=const.$ submanifold. Consider eqs. (\ref{al81}) and (\ref{al83}).  Once $I_{2}$ is fixed (and viewed as a parameter) they describe the Hamiltonian flow on twodimensional symplectic manifold parametrized by $\Theta_{1}$ and $I_{1}$ and equipped with the symplectic form $dI_{1}\wedge d\Theta_{1}$. The invariant symplectic structures are classified by Kirillov forms on coadjoint orbits. For $SU(2)$ such orbits are (apart from the trivial one) twodimensional spheres or, equivalently, the coset space $SU(2)/U(1)$. Guided by this observation we write out the general element $g\in SU(2)$ in the form
\begin {align} 
\label {al85}
g&=\left(\cos\left(\frac{\alpha}{2}\right)\mathbb{1}-i \sin\left(\frac{\alpha}{2}\right)\vec{n}\cdot\vec{\sigma}\right)e^{-i\tilde{\Theta}\frac{\sigma_{3}}{2}}\equiv\\ \nonumber
&\equiv \left(\cos\left(\frac{\alpha}{2}\right)\mathbb{1}-i \sin\left(\frac{\alpha}{2}\right)\vec{n}\cdot\vec{\sigma}\right)\left(\cos\left(\frac{\tilde{\Theta}}{2}\right)\mathbb{1}-i \sin \left(\frac{\tilde{\Theta}}{2}\right)\sigma_{3}\right)
\end {align}
where $\mathbb{1}$ and $\vec{\sigma}$ are the unit and Pauli matrices, respectively, while $\vec{n}$ is an unit vector in $12$ plane, 
\begin {align} 
\label {al86}
\vec{n}=&(\cos \delta, \sin \delta)\\ \nonumber
\text {and}\quad &0\leq\alpha\leq\pi\quad \text{,}\quad 0\leq\tilde{\Theta}<4\pi\quad\text{,}\quad 0\leq\delta<2\pi\;\text{.}
\end {align}
\par Let us consider the adjoint $SU(2)$ orbit passing through the element $\kappa\sigma_{3}$, $\kappa\in\mathbb{R}$ ($SU(2)$ is (semi)simple so the adjoint and coadjoint orbits are equivalent). It reads
\begin {align} 
\label {al87}
g(\kappa\sigma_{3})g^{-1}=B_{k}\sigma_{k}
\end {align}
where
\begin {align} 
\label {al88}
B_{1}&=\kappa\sin\alpha\sin\delta\\\nonumber
B_{2}&=-\kappa\sin\alpha\cos\delta\\\nonumber
B_{3}&=\kappa\cos\alpha
\end {align}
$B'_{i}s$ obey $sU(2)$ algebra (with respect to the Poisson bracket implied by Kirillov form) so one can identify them with $B'_{i}s$ given by eqs. (\ref{al70})-(\ref{al72}). Hence, we get the following relations
\begin {align} 
\label {al89}
I_{2}=\kappa
\end {align}
\begin {align} 
\label {al90}
\frac{I_{1}}{I_{2}}=\cos\alpha
\end {align}
\begin {align} 
\label {al91}
\Theta_{1}=\delta
\end {align}
\par It remains to relate $\Theta_{2}$ and $\tilde{\Theta}$. To this end we compare the infinitesimal transformation properties of both sets of variables. Those of $\Theta_{i}$, $I_{i}$, $i=1,2$, can be readily read off from eqs. (\ref{al81})-(\ref{al84}). On the other hand, let us find the transformation rules for $\alpha$, $\delta$ and $\tilde{\Theta}$ under left multiplication. We find
\begin {align} 
\label {al92}
\left(\mathbb{1}-\frac{i\delta\epsilon}{2}\sigma_{1}\right)g=\left(\cos\left(\frac{\alpha'}{2}\right)\mathbb{1}-i\sin\left(\frac{\alpha'}{2}\right)\vec{n}'\cdot\vec{\sigma}\right)e^{-\frac{i\tilde{\Theta}'\sigma_{3}}{2}}
\end {align}
with
\begin {align} 
\label {al93}
\delta\alpha=\delta\epsilon\cdot\cos\delta
\end {align}
\begin {align} 
\label {al94}
\delta\delta=-\delta\epsilon\frac{\cos\alpha}{\sin\alpha}\sin\delta
\end {align}
\begin {align} 
\label {al95}
\delta\tilde{\Theta}=+\delta\epsilon\frac{\sin\left(\frac{\alpha}{2}\right)}{\cos\left(\frac{\alpha}{2}\right)}\sin\delta
\end {align}
Taking into account the identification (\ref{al89})-(\ref{al91}) we find that the transformation properties (\ref{al93})-(\ref{al95}) coincide with those implied by (\ref{al81})-(\ref{al84}) provided we additionally identify
\begin {align} 
\label {al96}
\tilde{\Theta}=\Theta_{1}+\Theta_{2}
\end {align}
\par The same conclusions concern the symmetry transformations generated by $B_{2}$ and trivially $B_{3}$. So the symmetry under consideration may be described as left group multiplication of the suitably parametrized (in terms of original action-angle variables) $SU(2)$ group elements. However, one point should be clarified. Multiplication by the central element $g=-\mathbb{1}$ of $SU(2)$ corresponds, via (\ref{al85}) and (\ref{al96}), to the replacement $\Theta_{2}\to\Theta_{2}+2\pi$, i.e. corresponds to the same point of phase space. Therefore, it is rather $SO(3)$ than $SU(2)$ which action is generated by Runge-Lenz (suitably normalized) vector.
\section{Summary} 
\label{VI}
\par As for all static centrally symmetric potentials the energy and angular momentum are the integrals of motion. Their conservation follows, via Noether theorem, from the symmetries under time translations and space rotations which are point symmetries defined in configuration space. The Kepler problem exhibits additional conservation law. The conserved quantity is the Runge-Lenz vector. Its form (more precisely - momentum dependence) suggests that it cannot be derived from the symmetry described by point transformations. Therefore, one has to consider more general symmetries described by the canonical transformations. It is quite easy to find their infinitesimal form. The problem arises how to determine the global form. The Poisson brackets of angular momentum and Runge-Lenz vector form $SO(4)$ algebra on the submanifold of fixed negative energy. This suggests that the global symmetry transformations should describe the action of $SO(4)$ on the phase space. 
\par In order to simplify the problem we make use of the fact that the motion is plane. Therefore, the $SO(4)$ symmetry is "spontaneously" broken to $SO(3)$ which considerably simplifies (technically) the problem. The fully threedimensional case can be quite easily recovered because it is sufficient to add one generator of ordinary rotations (one component of angular momentum). 
\par First, we considered the canonical transformations generated by an arbitrary linear combination $\vec{c}\cdot\vec{A}$ of the components of Runge-Lenz vector. They are obtained from the Hamiltonian equations corresponding to the "Hamiltonian" $\vec{c}\cdot\vec{A}$. Their form is quite complicated. However, we were able to show that, using $sO(3)$ algebra spanned by $\vec{A}$ and $L$, the evolution equations can be reduced to a single first order nonlinear differential equation with variable coefficients. Such equations are usually not integrable by quadratures. However, our "Hamiltonian" system is Liouville integrable because it possesses two functionally independent Poisson commuting integrals: $H$ and $\vec{c}\cdot\vec{A}$. Therefore, it \underline{is} integrable by quadratures and the same applies to our final equation; the explicit form of the solution seems to be tremendously complicated. In order to simplify the problem we passed to the action-angle variables (defined relative to the initial Hamiltonian $H$). Then it appears that the solution is given explicitly, cf. eqs. (\ref{al55})-(\ref{al60}). Once we know it we can go back to initial variables $\vec{q} $, $\vec{p}$ provided we solve a transcendental equation of Kepler's type.\\
It appears that the generic trajectories are not closed which provides the evidence that they do not represent the action of $SO(3)$ group. The reason for that is explained in Sec. V. It is noticed there that the vector fields defining the infinitesimal action of symmetry transformations do not close to Lie algebra and Lie-Palais theorem is not applicable. This can be cured be redefining the Runge-Lenz vector; the conservation of the latter is equivalent to that of the normalized one. New conserved charges generate global $SO(3)$ transformations. They can be nicely described as follows. One writes the general element $g\in SU(2)$ in the form
\begin {align} 
\label {al97}
g=e^{-\frac{i\alpha}{2}\vec{n}\cdot\vec{\sigma}}e^{-\frac{i\tilde{\Theta}\sigma_{3}}{2}}
\end {align}
where
\begin {align} 
\label {al98}
\cos\alpha=\frac{I_{\varphi}}{I_{r}+\vert I_{\varphi}\vert}\qquad\text{,}\qquad 0\leq\alpha\leq\pi
\end {align}
\begin {align} 
\label {al99}
\vec{n}=\left(\cos\left(\Theta_{\varphi}-\Theta_{r}\right),\sin\left(\Theta_{\varphi}-\Theta_{r}\right),0\right)
\end {align}
\begin {align} 
\label {al100}
\tilde{\Theta}=\Theta_{\varphi}
\end {align}
with $\Theta_{r}$, $\Theta_{\varphi}$, $I_{r}$, $I_{\varphi}$ being the action-angle variables for the initial Kepler dynamics in polar coordinates. Then the symmetry transformations related, via Noether theorem in Hamiltonian form, to the conservation of normalized Runge-Lenz vector, are simply given by the left multiplication of $g$ by an arbitrary element of $SU(2)$.
\par Two remarks are in order:\\
\labelitemi\;replacing $\Theta_{\varphi}\to\Theta_{\varphi}+2\pi$ yields the same point in phase space; on the other hand, $g$, as given by eq. (\ref{al97}), goes into $-g$. Therefore, the global symmetry is $SO(3)$ rather than $SU(2)$;\\
\labelitemii\; eq. (\ref{al98}) implies that $I_{\varphi}$ can be negative. This is because we would like to keep the relation $I_{\varphi}=p_{\varphi}$ valid whether the particle moves clockwise or counterclockwise while the standard definition implies that the action variable is positive. Admitting $I_{\varphi}=p_{\varphi}$ (i.e. both signs of $I_{\varphi}$) is necessary if we insist on equations of motion (\ref{al81})-(\ref{al84}) to be valid beyond the region $I_{\varphi}\geq0$.\\
\labelitemiii\; $SU(2)$ acts on the phase space as the group of nonlinear transformations. The theory of such nonlinear realizations (especially elegant in the case of compact groups) has been developed in seminal papers \cite{b19}. Referring to their terminology the stability subgroup $\mathcal{H}=U(1)$ is generated by $\frac{1}{2}\sigma_{3}$; the preferred variables, parametrizing the coset space $SU(2)/U(1)$ are $\Theta_{1}$ and $I_{1}/I_{2}$ while the adjoint ones, $\Theta_{1}+\Theta_{2}$ and $I_{2}$, correspond to the character $1$ and $0$ representations of $\mathcal{H}=U(1)$, respectively. This provides the description of the action of symmetry group $SU(2)$ $(E<0)$ in terms of standard notions used in the theory of nonlinear realizations.

\appendix
\section*{Appendix\\Liouville integrability in Cartesian\\ coordinates}
\par Assume we are interested in canonical transformations generated by the $A_{1}$ component of Runge-Lenz vector. There are two Poisson-commuting integrals of motion
\begin {align} 
\label {al101}
E=\frac{\vec{p}\,^{2}}{2m}-\frac{k}{\vert\vec{q}\vert}\qquad (E<0)
\end {align}
\begin {align} 
\label {al102}
A_{1}=\vec{p}\,^{2}q_{1}-(\vec{p}\cdot\vec{q})p_{1}-\frac{mkq_{1}}{\vert\vec{q}\vert}
\end {align}
which may be rewritten as
\begin {align} 
\label {al103}
p^{2}_{1}+p^{2}_{2}=2mE+\frac{2mk}{\vert\vec{q}\vert}
\end {align}
\begin {align} 
\label {al104}
(-p^{2}_{1}+p^{2}_{2})q_{1}-2q_{2}p_{1}p_{2}=2A_{1}-2mEq_{1}\cdot
\end {align}
One concludes from (\ref{al103}) and (\ref{al104}) that the relevant submanifold is noncompact. Therefore, one can take directly $E$ and $A_{1}$ as the new momenta. Computing $p_{1}$, $p_{2}$ from (\ref{al103}), (\ref{al104}) one defines the generating function
\begin {align} 
\label {al105}
S=\int(p_{1}dq_{1}+p_{2}dq_{2})
\end {align}
\par Then the new coordinates are given by
\begin {align} 
\label {al106}
\Psi_{E}=\frac{\partial S}{\partial E}\qquad \text {,}\qquad \Psi_{A}=\frac{\partial S}{\partial A_{1}}
\end {align}
while the equations of motion read
\begin {align} 
\label {al107}
\frac{d\Psi_{E}}{d\epsilon}=0\qquad \text {,}\qquad \frac{d\Psi_{A}}{d\epsilon}=1
\end {align}
\par Therefore, the solution can be read off from
\begin {align} 
\label {al108}
\Psi_{H}(\vec{q},E,A_{1})&=const\\ \nonumber
\Psi_{A}(\vec{q},E,A_{1})&=\epsilon + const
\end {align}


\end {document}